\documentclass[lettersize,journal]{IEEEtran}

\usepackage{amsmath,amsfonts}
\usepackage{algorithmic}
\usepackage{algorithm}
\usepackage{array}
\usepackage[caption=false,font=normalsize,labelfont=sf,textfont=sf]{subfig}
\renewcommand{\baselinestretch}{0.86}
\usepackage{textcomp}
\usepackage{stfloats}
\usepackage{xcolor}
\usepackage{url}
\usepackage{verbatim}
\usepackage{graphicx}
\usepackage{cite}
\usepackage{booktabs}
\usepackage{listings}
\usepackage{xcolor}
\usepackage[colorlinks,linkcolor=black,anchorcolor=black,citecolor=black]{hyperref}
\usepackage{caption}

\begin{document}

\title{LLM-DMD: Large Language Model-based Power System \\Dynamic Model Discovery}

\author{Chao Shen,~\IEEEmembership{Student Member,~IEEE,}
Zihan Guo,~\IEEEmembership{Student Member,~IEEE,}
Ke Zuo,~\IEEEmembership{Student Member,~IEEE,}\\
Wenqi Huang,~\IEEEmembership{Member,~IEEE,}
Mingyang~Sun,~\IEEEmembership{Senior Member,~IEEE}
\thanks{
(\textit{Chao Shen and Zihan Guo are co-first authors.}
)



}
}


\maketitle

\begin{abstract}
\textcolor{black}{Current model structural discovery methods for power system dynamics impose rigid priors on the basis functions and variable sets of dynamic models while often neglecting algebraic constraints, thereby limiting the formulation of high-fidelity models required for precise simulation and analysis. This letter presents a novel large language model (LLM)-based framework for dynamic model discovery (LLM-DMD) which integrates the reasoning and code synthesis capabilities of LLMs to discover dynamic equations and enforce algebraic constraints through two sequential loops: the differential-equation loop that identifies state dynamics and associated variables, and the algebraic-equation loop that formulates algebraic constraints on the identified algebraic variables.
In each loop, executable skeletons of power system dynamic equations are generated by the LLM-based agent and evaluated via gradient-based optimizer. Candidate models are stored in an island-based archive to guide future iterations, and evaluation stagnation activates a variable extension mechanism that augments the model with missing algebraic or input variables, such as stator currents to refine the model. Validation on synchronous generator benchmarks of the IEEE 39-bus system demonstrates the superiority of LLM-DMD in complete dynamic model discovery.}





\end{abstract}


\vspace{-1em}
\section{Introduction}\label{Intro}

{\color{black}
\IEEEPARstart{T}{he} increasing complexity and stochasticity of modern power systems necessitate high-fidelity dynamic models for credible risk assessment and mitigation \cite{wang2020multi}. Model validation typically compares simulated responses with phasor measurement unit (PMU) data recorded during disturbances \cite{stankovic2020data}, where observed mismatches often trigger model refinement \cite{hu2022toward} through either parameter identification (PI) or structural discovery (SD). PI methods, ranging from gradient approach to deep reinforcement learning \cite{hu2022toward}, adjust parameters within a fixed model structure but cannot address structural deficiencies \cite{saric2020symbolic}. In contrast, SD frameworks attempt to reconstruct both the governing models and their parameters from data \cite{saric2020symbolic}. 

Despite these advances, discovering the complete differential algebraic equations (DAEs) for power system dynamics requires substantial domain expertise, since the symbolic search space grows combinatorially with uncertainties in basis functions \cite{shojaee2024llm} and auxiliary variables beyond the state vector \cite{holt2024data}. A representative SD approach is the sparse identification of nonlinear dynamics (SINDy) \cite{stankovic2020data,saric2020symbolic}, which employs sparsity assumptions promoting symbolic regression paradigms. Specifically, SINDy-style pipelines mitigate these uncertainties by imposing rigid expert priors, fixing the variable set, and restricting basis functions to polynomial and trigonometric dictionaries \cite{saric2020symbolic, stankovic2020data}. However, such accurate priors are seldom attainable \cite{holt2024data}, and their misspecification can result in structurally biased or even unidentifiable dynamic models. Moreover, existing pipelines in \cite{stankovic2020data,saric2020symbolic} primarily target differential equations without a systematic identification of algebraic constraints, resulting in incomplete DAEs that limit their applicability to system-wide dynamic studies.}

To address these challenges, this letter proposes a novel large language model-based dynamic model discovery framework (LLM-DMD), with the contributions summarized below:

1) To the best of our knowledge, this is the first SD framework for power system dynamics that integrates LLM-based reasoning and code synthesis with gradient-based evaluation, island-based feedback, and dynamic variable expansion to enable joint identification of governing differential equations, algebraic constraints and parameters.

2) The proposed framework decouples discovery into a differential-equation (DE) loop that formulates state dynamics and identifies the involved variables, and an algebraic-equation (AE) loop that imposes constraints on the algebraic variables revealed by the DE loop.

3) An LLM-based modeling agent (M-Agent) is designed to generate executable skeletons of power system DAEs guided by modeling task contracts and feedback in-context examples, facilitating structured synthesis of state equations and algebraic constraints without predefined basis functions.

4) A variable extension mechanism is developed, where evaluation stagnation triggers the parsing of JSON-formatted requirements to adaptively introduce algebraic/input variables, such as stator currents, for capturing unmodeled dependencies.

\vspace{-1em}
{\color{black}
\section{Problem Formulation}
Power-system dynamics are typically represented by DAEs:
\begin{align}
\dot{\boldsymbol{x}}(t) &= \boldsymbol{f}(\boldsymbol{x}(t), \boldsymbol{y}(t), \boldsymbol{u}(t), \boldsymbol{p}), \label{eq:diff}\\
\mathbf{0} &= \boldsymbol{g}(\boldsymbol{x}(t), \boldsymbol{y}(t), \boldsymbol{u}(t), \boldsymbol{p}). \label{eq:alg}
\end{align}
Here, $\boldsymbol{x}(t)\!\in\!\mathbb{R}^{n_x}$ denotes differential states (e.g., rotor angle $\delta$), $\boldsymbol{y}(t)\!\in\!\mathbb{R}^{n_y}$ represents algebraic variables (e.g., bus voltage magnitude $V$ and phase $\theta$), $\boldsymbol{u}(t)\!\in\!\mathbb{R}^{n_u}$ is inputs (e.g., mechanical power $P_m$), and $\boldsymbol{p}\!\in\!\mathbb{R}^{n_p}$ is parameters.} SD aims to estimate ${\boldsymbol{{f}}}(\cdot)$ and ${\boldsymbol{{g}}}(\cdot)$ from a dataset $\mathcal{D}=\{(\boldsymbol{x}_i,\dot{\boldsymbol{x}}_i)\}_{i=1}^{n_s}$ \cite{saric2020symbolic,stankovic2020data,shojaee2024llm} where $\dot{\boldsymbol{x}}_i$ obtained by numerical differentiation. The task is ill-posed because, {prior to discovery}, neither (i) the involved algebraic/input set $\mathcal V$ (i.e., which components of $\boldsymbol{y}$ and $\boldsymbol{u}$ enter the equations) nor (ii) the appropriate candidate function libraries for $\boldsymbol f$ and $\boldsymbol g$ (e.g., polynomial or trigonometric) is known a priori. 


\vspace{-1em}
\section{Methodology}
\subsection{An Overview of Proposed LLM-DMD Framework}

\begin{figure}[H]
  \centering
  \includegraphics[width=0.85\linewidth]{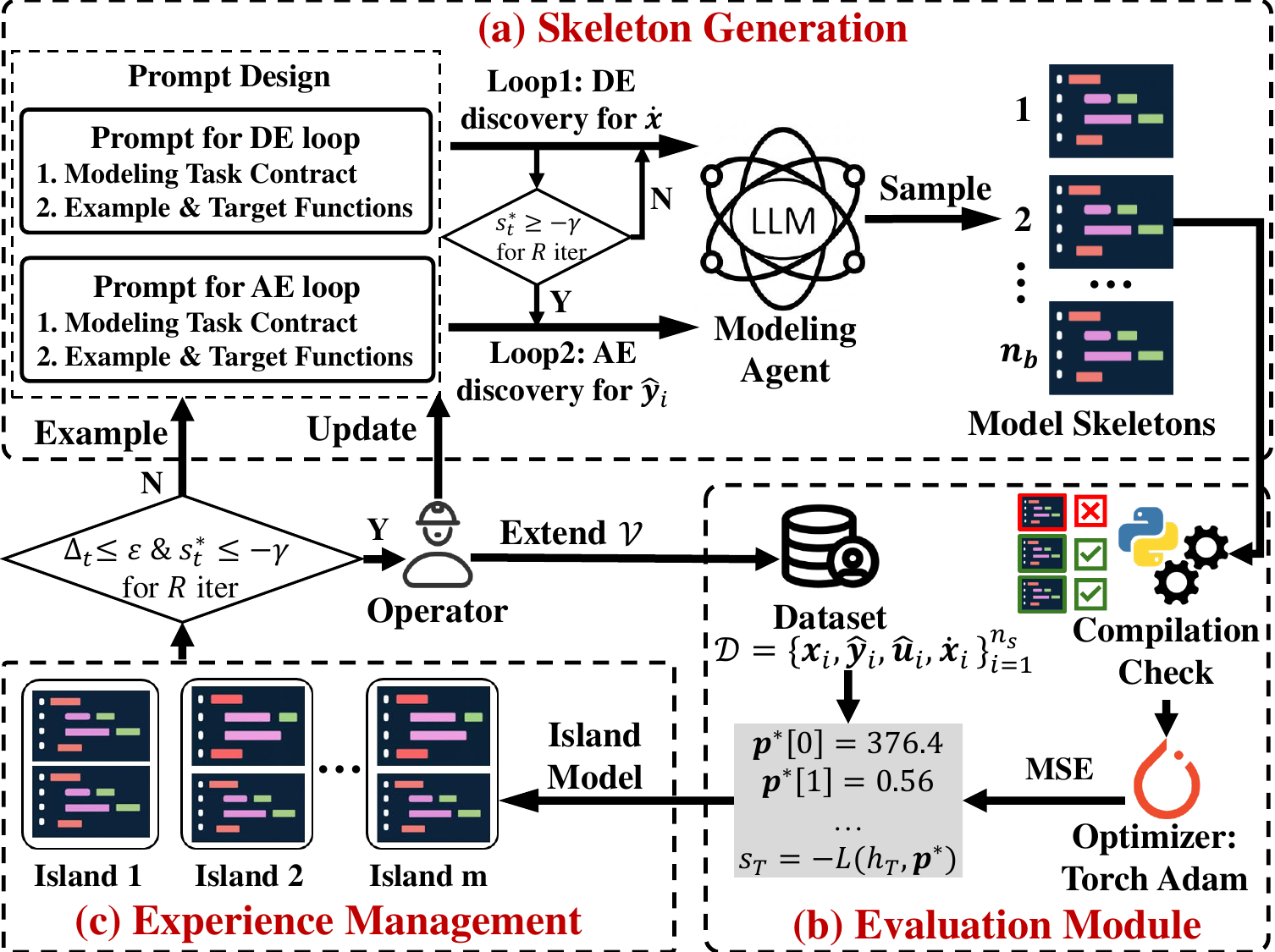}
  \caption{Framework for LLM-DMD.}
  \label{Framework for LLM-DMD}
\end{figure}

As shown in \autoref{Framework for LLM-DMD}(a), the proposed LLM-DMD framework develops two sequential loops to identify differential state equations and algebraic constraints. In the DE loop, the LLM-based M-Agent generates executable DE code skeletons ${\boldsymbol{\hat{f}}}(\cdot)$ under a DE-specific prompt. Non-compilable candidates are discarded, while valid ones are evaluated through gradient-based optimization (\autoref{Framework for LLM-DMD}(b)) and archived in an experience island for subsequent example selection (\autoref{Framework for LLM-DMD}(c)). When convergence stagnates, the involved variable set $\mathcal{V}$ is adaptively expanded, and iterations continue until the differential equations ${\boldsymbol{\hat{f}}}(\boldsymbol{x},\hat{\boldsymbol{y}},\hat{\boldsymbol{u}},\boldsymbol{p})$ and involved variables $\mathcal{V}=\{\boldsymbol{\hat{y}},\boldsymbol{\hat{u}}\}$ are identified. {\color{black}With $\hat{\boldsymbol y}$ fixed by the DE loop, the number of algebraic constraints can be inferred and the AE loop discovers an explicit mapping $\hat{\boldsymbol y}=\hat{\boldsymbol g}(\boldsymbol x,\hat{\boldsymbol u},\boldsymbol p)$ (the implicit form $\mathbf{0}=\boldsymbol g(\boldsymbol x,\boldsymbol {\hat{y}},\boldsymbol {\hat{u}},\boldsymbol p)$ can be handled analogously), aligning with the DE loop and enabling reuse of the same generate–evaluate–feedback pipeline under AE-specific prompts. 

}

\vspace{-1em}

\vspace{-2em}
\begin{figure}[H]
  \centering
  \includegraphics[width=0.9\linewidth]{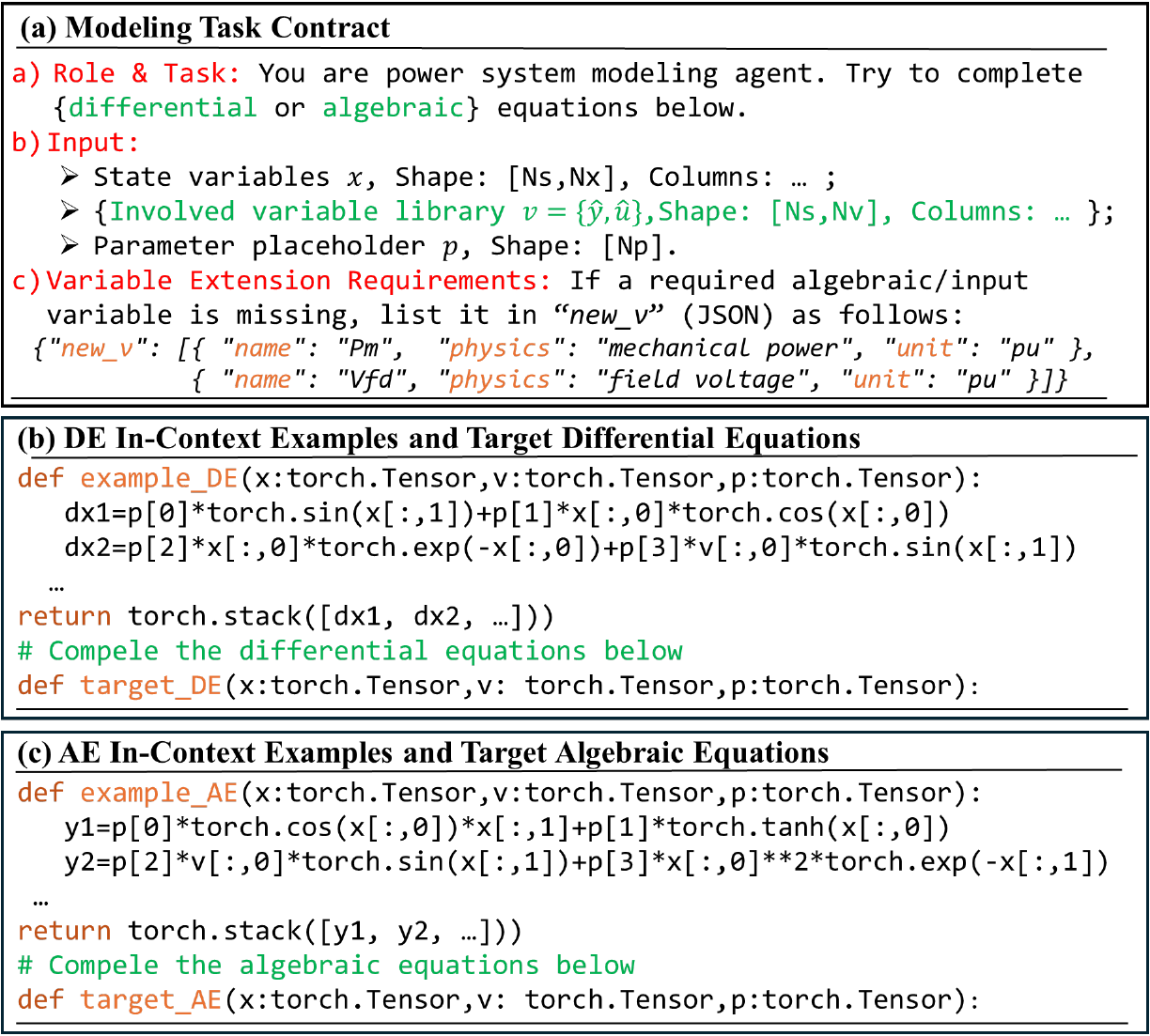}
  \caption{Prompt design for M-Agent.}
  \label{Prompt Design for Modeling Agent}
\end{figure}

\vspace{-2em}
\subsection{Model Skeleton Generation with M-Agent Prompting}
{\color{black}As shown in \autoref{Framework for LLM-DMD}(a), the M-Agent synthesizes executable model skeletons for DEs and AEs by leveraging inherent reasoning and code synthesis capability. For each loop $T \in \{DE,AE\}$, the input prompt $\pi_T$ comprises a task contract (\autoref{Prompt Design for Modeling Agent}(a)) and in-context examples with a target function (\autoref{Prompt Design for Modeling Agent}(b)(c)). The contract (\autoref{Prompt Design for Modeling Agent}(a)) specifies the modeling role, equation completion rules, and variable and parameter descriptions. In the DE loop, the variable library $\mathcal{V}$ is initialized as empty and expanded adaptively based on the structured JSON requirements from M-Agent. Once the DE loop converges, the identified algebraic variables $\hat{\boldsymbol{y}}$ is fixed from ${\boldsymbol{\hat{f}}}(\boldsymbol{x},\hat{\boldsymbol{y}},\hat{\boldsymbol{u}},\boldsymbol{p})$ to initialize the AE loop. The AE contract differs from the DE version in task description and variable library, which is inherited from the final $\mathcal{V}$ of the DE loop, denoted by "$\{\cdot\}$" and green annotations in \autoref{Prompt Design for Modeling Agent}(a). Its examples and target function are adapted to emphasize the explicit representation of $\hat{\boldsymbol{y}}$ (\autoref{Prompt Design for Modeling Agent}(c)).
}

{\color{black}
Given $\pi_T$ and temperature $\tau$, $n_b$ skeletons are sampled from the softmax distribution induced by M-Agent, $\mathbb{P}_T(\cdot \mid \pi_T, \tau)$:
\begin{equation}
  \widetilde{\mathcal{H}}_{T} = \big\{ h_{T}^{(j)} \sim \mathbb{P}_{T}(\cdot \mid \pi_{T}, \tau) \big\}_{j=1}^{n_b}, ~ T\in \{DE,AE\}.
\end{equation}
where $h^{(j)}_T$ denotes the $j$-th generated candidates for $\boldsymbol{\hat{f}}$ or $\boldsymbol{\hat{g}}$. Each skeleton substitutes numeric parameters with trainable placeholders $\texttt{p}$ (\autoref{Prompt Design for Modeling Agent}(b),(c)). Non-executable candidates are discarded by compilation, and the accepted set is defined as $\mathcal{H}_{T}=\{\, h_t \in \widetilde{\mathcal{H}}_{T} \mid \chi(h_T)=1 \,\}$ where $\chi(h_T)=1$ if compilation succeeds and $\chi(h_T)=0$ otherwise.
}

\vspace{-1em}
\subsection{Parameter Estimation and Skeleton Evaluation}
{\color{black}
For each candidate skeletons $h_T^{(j)} \in \mathcal{H}_T$, the evaluation module (\autoref{Framework for LLM-DMD}(b)) fits parameters $\boldsymbol{p}$ by minimizing the mean squared error with loop-specific losses
\begin{align}
L_{\mathrm{DE}}(\boldsymbol{p}) 
&= \frac{1}{n_s} \sum_{i \in \mathcal{D}} 
   \big\| h_{\mathrm{DE}}^{(j)}(\boldsymbol{x}_i,\hat{\boldsymbol{y}}_i^{DE},\hat{\boldsymbol{u}}_i^{DE},\boldsymbol{p}) 
      - \dot{\boldsymbol{x}}_i \big\|_2^2, \\
L_{\mathrm{AE}}(\boldsymbol{p}) 
&= \frac{1}{n_s} \sum_{i \in \mathcal{D}} 
   \big\| h_{\mathrm{AE}}^{(j)}(\boldsymbol{x}_i,\hat{\boldsymbol{y}}_i^{AE},\hat{\boldsymbol{u}}_i^{AE},\boldsymbol{p}) 
      - \hat{\boldsymbol{y}}_i^{DE} \big\|_2^2 .
\end{align}
Here, $\mathcal{V}^T=\{\hat{\boldsymbol{y}}^T,\hat{\boldsymbol{u}}^T\}$ ($T\in\{DE,AE\}$) denotes the loop-specific algebraic and input variables. In $L_{\mathrm{AE}}$, the DE-loop identified $\hat{\boldsymbol{y}}^{DE}_i$ serve as labels to identify parameters within $h_{\mathrm{AE}}^{(j)}$.
Parameters are optimized with Adam using a cosine-annealed learning rate, yielding
$\boldsymbol{p}^*$, and the skeleton score is
\begin{align}
s(h_T^{(j)}) = -\,L_T(\boldsymbol{p}^*), \quad T \in \{DE,AE\}.
\end{align}
Each skeleton is thus assigned the score $\{h_T^{(j)},s\}$ and forwarded to the experience management module.

}

\vspace{-1em}
\subsection{Feedback and Variables Extension}
\subsubsection{Experience Management with Island-Based Evolution}
{\color{black}The experience management module (\autoref{Framework for LLM-DMD}(c)) utilizes an island-based evolutionary strategy to maintain diversity throughout the discovery process and sample promising model skeletons that guide subsequent iterations. It operates with $m$ islands, $\{\mathcal{I}_k\}_{k=1}^m$, each initialized with a seed skeleton of $\boldsymbol{\hat{f}}$ or $\boldsymbol{\hat{g}}$ in linear form. These islands are progressively enriched by evaluating candidates $\{h_T^{(j)},s\}$, which are then grouped into clusters, $\{C_{k,c}\}_{c=1}^{n_c}$, based on their score similarity.

In-context examples as detailed in \autoref{Prompt Design for Modeling Agent}(b)(c) are sampled as follows: 1) An island $k \in \{1, \dots, m\}$ is selected uniformly to ensure equal exploration; 2) A cluster $C_{k,c}$ and a skeleton $h_T \in C_{k,c}$ are drawn sequentially by the softmax distribution:
\begin{equation}
    \mathbb{P}(z) = \frac{\exp(\phi(z)/\tau)}{\sum_{z'} \exp(\phi(z')/\tau)},
\end{equation}
where $z$ represents either a cluster or a skeleton, and $\tau$ controls the sampling randomness. For cluster selection, $\phi(C_{k,c}) = s
_{k,c}$ denotes the mean score of cluster $C_{k,c}$, with temperature $\tau_c$; for skeleton selection, $\phi(h_T) = -\ell(h_T)$, where $\ell(h_T)$ is the code length, encouraging concise and high-scoring formulations with temperature $\tau_\ell$.
}

\subsubsection{Variable Extension Mechanism} 
{\color{black}Before each iteration, the framework monitors progress by the global best score across all islands:
\begin{align}
  s^*_t=\max_{h_T\in \{I_k\}_{k=1}^m} s(h_T).
\end{align}
where $t$ is the iteration index. The incremental gain is given by $\Delta_t = s_t^* - s_{t-1}^*$. The variable extension mechanism is triggered when the search stagnates for $R$ consecutive iterations, i.e.,
\begin{equation}
\max_{t \in {t-R+1,\ldots,t}} \Delta_t \le \varepsilon,~ \text{and} ~ s_t^* \le -\gamma,
\label{eq:trigger}
\end{equation}
with $\varepsilon$ and $\gamma$ denoting small positive thresholds ($\gamma\le \varepsilon$). Once triggered, JSON-formatted requirements (\autoref{Prompt Design for Modeling Agent}(a)) are extracted from the top-$k$ skeletons to identify new algebraic and input variables proposed by the M-Agent. The dataset is then augmented via measurement or simulation, and the variable library $\mathcal{V}$ in the prompt contract is updated for the next iteration. The loops terminate when $s^*_t$ remains above $-\gamma$ for $R$ consecutive iterations.}

\section{Case Study}
\subsection{Experiment Setup}
{\color{black}The framework is validated on the IEEE 39-bus, 10-machine system. Load levels vary within $\pm$30\% of nominal, and operating points are obtained via optimal power flow. The datasets are generated by applying three-phase-to-ground faults at bus 15 (training) and bus 28 (testing) and running 10-s time-domain simulations with a 0.01-s step in PSAT \cite{milano2005open}. 
Gaussian noise ($\sigma=1\%$ of amplitude) is added on the state trajectories of training set to emulate measurement errors. The modeling target is the dynamics of synchronous generators (SGs) \cite{stankovic2020data,saric2020symbolic} with the ground truth given by the fifth-order, type \uppercase\expandafter{\romannumeral 1} model detailed in \cite{milano2005open, Milano2006PSATManual}. The states consist of rotor angle $\delta$, rotor speed $\omega$, $q$-axis transient voltage $e_q^{\prime}$, and $d$-axis transient and subtransient voltages $e_d^{\prime}, e_d^{\prime\prime}$. The algebraic variables are stator currents $i_d,i_q$ and electromagnetic power $P_e$, while the inputs include mechanical power $P_m$ and excitation voltage $v_f^*$. The framework is implemented in Python/PyTorch with the M-Agent built on DeepSeek-V3 \cite{deepseekai2024deepseekv3technicalreport}, with hyperparameters $m=10$, $\tau=1.2$, $\tau_c=\tau_{\ell}=0.2$, $\varepsilon=\gamma=0.01$, and $R=3$.}

\vspace{-2em}
\begin{figure}[H]
  \centering
  \includegraphics[width=0.8\linewidth]{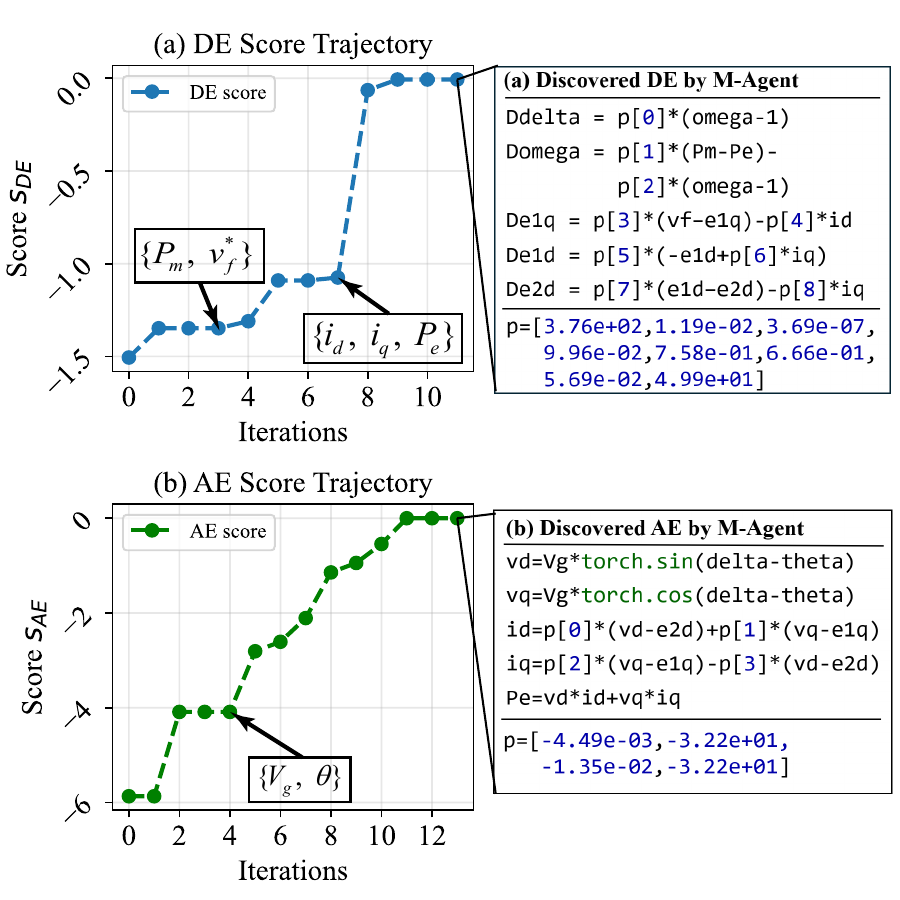}
  \vspace{-1em}
  \caption{$s^*_t$ trajectories in the DE/AE loops for SG at bus 31.}
  \label{Score trajectory in DE and AE loop for SG at bus 31}
\end{figure}
\vspace{-2em}
\subsection{SG Dynamic Model Discovery via LLM-DMD Framework}
{\color{black}
The discovery process of the LLM-DMD is demonstrated on the SG at bus 31. The DE and AE are recovered within 11 and 13 iterations, respectively, with $s^*_t$ trajectories and the discovered models shown in \autoref{Score trajectory in DE and AE loop for SG at bus 31}. In the DE loop, the M-Agent starts with empty $\mathcal{V}$, and the score stagnates around $-1.5$ (\autoref{Score trajectory in DE and AE loop for SG at bus 31}(a)). Incorporating inputs $P_m$ and $v_f^*$ raises $s^*$ to about $-1.2$. A pronounced gain occurs once the stator currents $i_d, i_q$ and electromagnetic power $P_e$ are promoted to required variables, yielding a sharp ascent of $s^*$ toward zero and convergence to the accurate fifth-order DEs. In the subsequent AE loop, $i_d, i_q$, and $P_e$ are treated as targeted algebraic variables while $P_m$ and $v_f^*$ 
remain exogenous inputs from the prime mover and AVR, respectively, and are therefore excluded from discovery. As shown in \autoref{Score trajectory in DE and AE loop for SG at bus 31}(b), introducing the terminal-voltage magnitude $V_g$ and angle $\theta$ further reduces the discrepancy, lifting the score from $-4$ to nearly $-2$. Iterative refinement then drives the error to negligible levels, thereby recovering the algebraic constraints. Collectively, LLM-driven skeleton generation, parameter estimation, experience archiving, and variable extension resolve missing basis functions and latent variables, enabling near-autonomous identification of SG dynamics with operator effort limited to data provision.}

\subsection{Comparison with the SINDy Baseline}
{\color{black}
{\color{black}
We benchmark the proposed framework against SINDy \cite{saric2020symbolic,stankovic2020data} for DE discovery under controlled prior-knowledge conditions, consistent with prior studies that target DE discovery rather than full DAE recovery. The baseline \textit{SINDy‑Accurate‑Priors} uses a first-order polynomial library over the complete variable set (accurate priors). Two stress conditions are defined: \textit{SINDy‑Overcomplete‑Library}, which appends quadratic terms to emulate an overcomplete library, and \textit{SINDy‑Missing‑Variables}, which removes a subset of variables to emulate missing-variable priors. Across 10 SGs, each identified model is simulated on a test set at identical operating points. Trajectory fit to ground-truth states is quantified by mean absolute percentage error (MAPE) and the coefficient of determination $R^2$. Lower MAPE and $R^2$ closer to unity indicate better agreement.}
Results summarized in \autoref{tab:model_comparison} show that even with accurate priors, baseline SINDy yields MAPE=$4.72\%$, $R^2=0.59$, underscoring the challenge of recovering full dynamics via symbolic regression. Under overextended libraries or reduced variable sets, performance degrades markedly (MAPE $>15\%$, $R^2<0$), revealing strong prior sensitivity and limited scalability. In contrast, the proposed framework achieves a MAPE of $0.22\%$ and an $R^2$ of $0.97$, indicating that LLM-guided model priors—implicitly encoded rather than explicitly prescribed—enable accurate and generalizable DAE discovery.
}

\begin{table}[H]
\centering
\caption{Comparison with SINDy-based Methods}
\begin{tabular}{ccc}
\hline
Models & MAPE & $R^2$ \\
\hline
SINDy‑Accurate‑Priors \cite{stankovic2020data,saric2020symbolic} & $4.72\% $ & $0.59$ \\
SINDy‑Overcomplete‑Library & $15.85\% $ & $-2.57$ \\
SINDy‑Missing‑Variables & $17.73\% $ & $-6.43 $ \\
Proposed LLM-DMD & $\textbf{0.22\%}$ & $\textbf{0.97}$ \\
\hline
\end{tabular}
\label{tab:model_comparison}
\end{table}


\vspace{-1em}
\section{Conclusion}
This letter proposes an LLM-based framework that discovers power system dynamic models through a DE loop for state dynamics and a subsequent AE loop for algebraic constraints. By combining LLM-generated skeletons, parameter estimation, experience archiving, and variable extension, the framework achieves accurate discovery on IEEE 39-bus SG benchmarks, with potential extensions to dynamic models of other power system components, such as grid-following/forming converters and virtual SGs.


\vspace{-1em}

\bibliographystyle{ieeetr}  
\bibliography{ref}

\end{document}